\documentclass[sigconf]{nimeart}
\usepackage[font=footnotesize,
            justification=centering]{subcaption}
\AtBeginDocument{%
  }

\setcopyright{cc}
\copyrightyear{2025}
\acmYear{2025}
\acmDOI{}
\acmConference[NIME '25]{International Conference on New Interfaces for Musical Expression}{June 24--27,2025}{Canberra, Australia}
\acmISBN{}

\begin{document}

\title{AI Harmonizer: Expanding Vocal Expression with a Generative Neurosymbolic Music AI System}

\author{Lancelot Blanchard}
\email{lancelot@media.mit.edu}
\orcid{0000-0003-1580-3116}
\affiliation{%
  \institution{MIT Media Lab}
  \city{Cambridge}
  \state{MA}
  \country{USA}
}

\author{Cameron Holt}
\email{camholt@mit.edu}
\affiliation{%
  \institution{Massachusetts Institute of Technology}
  \city{Cambridge}
  \state{MA}
  \country{USA}
}

\author{Joseph A. Paradiso}
\email{joep@media.mit.edu}
\affiliation{%
  \institution{MIT Media Lab}
  \city{Cambridge}
  \state{MA}
  \country{USA}
}

\renewcommand{\shortauthors}{Blanchard et al.}

\begin{abstract}
  Vocals harmonizers are powerful tools to help solo vocalists enrich their melodies with harmonically supportive voices. These tools exist in various forms, from commercially available pedals and software to custom-built systems, each employing different methods to generate harmonies. Traditional harmonizers often require users to manually specify a key or tonal center, while others allow pitch selection via an external keyboard–both approaches demanding some degree of musical expertise. The AI Harmonizer introduces a novel approach by autonomously generating musically coherent four-part harmonies without requiring prior harmonic input from the user. By integrating state-of-the-art generative AI techniques for pitch detection and voice modeling with custom-trained symbolic music models, our system arranges any vocal melody into rich choral textures. In this paper, we present our methods, explore potential applications in performance and composition, and discuss future directions for real-time implementations. While our system currently operates offline, we believe it represents a significant step toward AI-assisted vocal performance and expressive musical augmentation. We release our implementation on GitHub.\footnote{Our implementation is available at \url{https://github.com/mitmedialab/ai-harmonizer-nime2025}.}
\end{abstract}


\maketitle

\section{Introduction \& Previous Work}

Vocal harmonizers have long been a valuable tool for vocalists, enabling real-time harmonization and multi-voice effects. Over the years, various hardware solutions have been developed, with TC-Helicon leading the commercial market in harmonization pedals. These pedals generally require the user to set the key in which the system can generate notes, either manually or through another audio input (e.g., with a guitar playing chords). Vocal harmonizers have also been developed as part of research projects. A notable example is Jacob Collier's vocal harmonizer, developed by Ben Bloomberg \cite{bloomberg_making_2020}, which allows him to use a keyboard to decide the harmonic texture of the output and provide a vocal melody as input. In parallel, the automation of melodic harmonization, which alleviates the need for a keyboard or a manual setting of a key, has been explored in numerous ways, with methods such as probabilistic modeling \cite{paiement_probabilistic_2006, makris_probabilistic_2015}, dynamic programming \cite{yi_accomontage2_2022}, and weighted pitch context vectors \cite{kranenburg_algorithmic_2023}.

In recent years, machine learning has driven significant advancement in generative music. In melody harmonization, projects such as Google's CoCoNet \cite{huang_counterpoint_2017} and the Blob Opera \cite{li_david_blob_2020}  have demonstrated the potential of deep learning for polyphonic music generation, alongside many other systems \cite{lim_chord_2017, tsushima_function-_2017, yan_part-invariant_2018, sun_melody_2021, chen_surprisenet_2021, takahashi_emotion-driven_2022}. The application of Transformer-based models--originally designed for natural language processing--to symbolic music generation \cite{huang_music_2018, thickstun_anticipatory_2023} has opened new possibilities for automatic accompaniment and real-time applications \cite{wu_adaptive_2024, zhou_local_2024, blanchard_developing_2024}. Additionally, neural networks have shown remarkable success in voice synthesis, with architectures such as VITS \cite{kim_conditional_2021} and HuBERT \cite{10.1109/TASLP.2021.3122291} enabling the efficient manipulation of vocal timbre and style.

However, and to the best of our knowledge, no end-to-end Machine Learning-based system has been proposed for automatic vocal harmonization. Through this work, we propose our architecture, enabling the harmonization of any input vocal line as a four-part harmony chorale, and discuss our results.

\section{Methodology}

Our approach for automatic voice harmonization uses a few different existing models, and adds some important logic to connect them together and create a powerful end-to-end framework. We make use of three different model architectures:

\begin{itemize}
    \item Basic Pitch \cite{2022_BittnerBRME_LightweightNoteTranscription_ICASSP}, a model developed and trained by Spotify that can perform automatic music transcription;
    \item Anticipatory Music Transformer (AMT) \cite{thickstun_anticipatory_2023}, a variant of the Music Transformer architecture \cite{huang_music_2018} that enables better compositions through anticipation mechanisms; and 
    \item Retrieval-based Voice Conversion (RVC)\footnote{RVC is available at \url{https://github.com/RVC-Project/Retrieval-based-Voice-Conversion}.}, a model for singing voice conversion based on the VITS architecture that provides a toolset for fast and accurate singing voice synthesis with pitch and speaker conditioning.
\end{itemize}

We combine these architectures to accomplish four sequential tasks:

\begin{enumerate}
    \item First, we use Basic Pitch to convert our vocal melody to MIDI;
    \item Then, we use a custom-trained AMT model to generate a four-part harmony based on our input melody;
    \item We then extract the fundamental frequency ($f_0$) information from our vocal melody and shift it to fit our three new parts;
    \item Finally, we use RVC to synthesize these three new vocal lines and add them to our input melody.
\end{enumerate}

\begin{figure*}[ht!]
  \centering
  \includegraphics[width=0.7\linewidth]{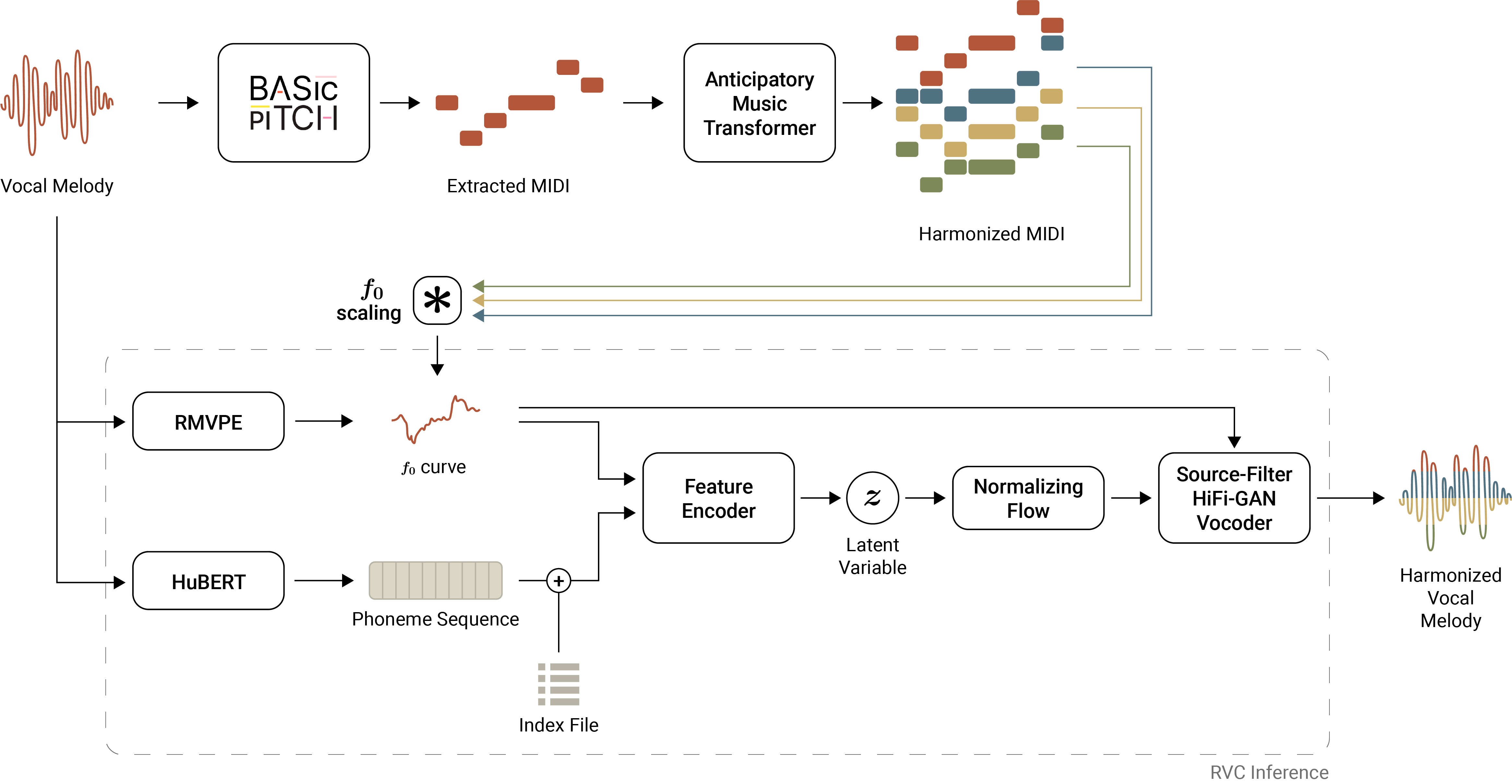}
  \caption{The detailed architecture of our system at inference time.}
  \label{fig:arch}
\end{figure*}

An overview of our architecture is displayed on Figure \ref{fig:arch}. We describe each step of our process in this section.

\subsection{Voice-to-MIDI Conversion}

To integrate our vocal melody into our custom-trained AMT model, we first need to convert the vocal line into MIDI. This is a relatively straightforward step, accomplished using \textit{Basic Pitch}, an open-source pitch detection tool developed by Spotify. While numerous pitch-tracking tools exist for this purpose, we chose Basic Pitch for its robustness in handling real-world singing inputs and its superior transcription accuracy. Extracting MIDI from vocal melodies is particularly challenging, as human singing often deviates from the strict 12-tone equal temperament required for MIDI representation. In our experiments, Basic Pitch appeared to handle vocal inputs fairly well. It is important to note that this approach comes with inherent limitations, as it restricts harmonization to Western tonal frameworks and favors cleanly sung inputs.

\subsection{Harmony Generation with AMT}

Once the vocal melody is transcribed into MIDI, we use a custom-trained Anticipatory Music Transformer model (AMT) to generate a four-part harmony that will be used to synthesize new vocal lines. AMT is particularly well-suited for this type of harmonic composition since it can \textit{anticipate} future notes in the melody and generate more thoroughly crafted harmonic lines.

\subsubsection{AMT Training}

In order to generate accurate harmonies for our vocal melody, we specifically train an AMT model on the task of four-part harmony. For this purpose, we base our training on the pre-trained \texttt{music-medium-800k}\footnote{The original checkpoint is available on Huggingface at \url{https://huggingface.co/stanford-crfm/music-medium-800k}.} model trained by the authors of the original paper on the Lakh MIDI Dataset \cite{raffel_learning-based_2016} for 800,000 epochs. We then choose to fine-tune our model on the \textbf{JSB Chorales} dataset \cite{NIPS2004_b628386c, boulanger-lewandowski_modeling_2012}, a corpus of 382 four-part harmonized chorales by J.S. Bach. Although this imposes an important genre restriction on the harmonies that our harmonizer can generate, this dataset is particularly interesting since all of the chorales are written with four distinct voices: \textit{Soprano}, \textit{Alto}, \textit{Tenor}, and \textit{Bass} (SATB). For the purpose of our training, we convert the original dataset to distinct MIDI files, and use MIDI instruments \texttt{0}, \texttt{1}, \texttt{2}, and \texttt{3} to represent all four voices.\footnote{We provide our version of the dataset on GitHub: \anon[https://anonymised/]{\url{https://github.com/lancelotblanchard/JSB-Chorales-dataset-midi}}.} Due to the fact that the data of the JSB Chorales dataset is most likely already contained in the Lakh MIDI Dataset, the model quickly overfits and we use early stopping to avoid model degradation.

\subsubsection{AMT Inference}

Anticipatory Music Transformers introduce the mechanism of \textit{anticipation}, which allows for the conditioning of a temporal point process on the realizations of another correlated process. Following the AMT naming convention, the main temporal point process is called the \textit{event} process, while the conditioning process is called the \textit{control} process. To perform this conditioning, AMT interleaves events $\textbf{e}_{1:N}$ and controls $\textbf{u}_{1:K}$ in such a way that a control $\textbf{u}_k$ on time $s_k$ ends up close to events near time $s_k-\delta$, with $\delta$ being the \textit{anticipation interval}. Using the results of the original paper, we use $\delta=5 \text{ seconds}$.

AMT's tokenization of MIDI notes allows us to precisely control the model generation. In particular, the model encodes a MIDI note as a triplet of time, duration, and note $(\textbf{t}_i,\textbf{d}_i,\textbf{n}_i)$, which allows us to guide the model generation to ensure that it can generate a well-founded four-part harmony. To do so, we enforce that each voice (MIDI instruments \texttt{1}, \texttt{2}, and \texttt{3}) can only generate \textit{one} harmony note for each note present in the input melody. We also ensure that the onset time and duration of each note overlap, by forcing the time token $\textbf{t}_i$ and duration token $\textbf{d}_i$ to take the value of the time and duration of the corresponding control. We do so by manually selecting the time and duration tokens, and by performing sampling for the note token on a restrained logit distribution, with the logits for notes from other instruments set to $-\infty$.

\subsection{MIDI-to-Frequency Conversion}

Once we have the MIDI information for our \textit{Alto}, \textit{Tenor}, and \textit{Bass} lines, we can start generating new vocal lines. To do so, we first need to extract the pitch contour (or fundamental frequency $f_0$) of our input melody. Although this information is also present in the MIDI data, the $f_0$ is a much more fine-grained measure that also contains the pitch fluctuations within notes, as well as the transition between notes. RVC provides a selection of algorithms and models for the purpose of pitch extraction. Among those, we choose to use \textit{RMVPE} \cite{wei_rmvpe_2023} for its robustness and execution speed.

Once the $f_0$ information is extracted from our original audio, we can start shifting it to match the pitch contour of our three harmony voices. Our approach for this task is straightforward: We first detect the onset of each note and delimit our $f_0$ curve based on those points, then shift the input curve.

Formally, given an input curve $f_0^{in}$ and a sequence of $N$ MIDI notes $\textbf{e}_{1:N}$ with $\textbf{e}_i = (\textbf{t}_i, \textbf{h}_i)$ (where we simplify the original AMT notation and consider that $\textbf{t}_i$ and $\textbf{h}_i$ respectively refer to the onset time of the note and the difference in semitones between the input note and the harmony voice), we have:

\[
f_0^{out} =
\begin{cases}
f_0^{in}(t) \cdot 2^{\textbf{h}_1/12}, &t_1 \leq t < t_2 \\
f_0^{in}(t) \cdot 2^{\textbf{h}_2/12}, &t_2 \leq t < t_3 \\
\vdots \\
f_0^{in}(t) \cdot 2^{\textbf{h}_N/12}, &t_N \leq t
\end{cases}
\]

\begin{figure}

\begin{subfigure}[t]{0.45\linewidth}
    \includegraphics[width=\textwidth]{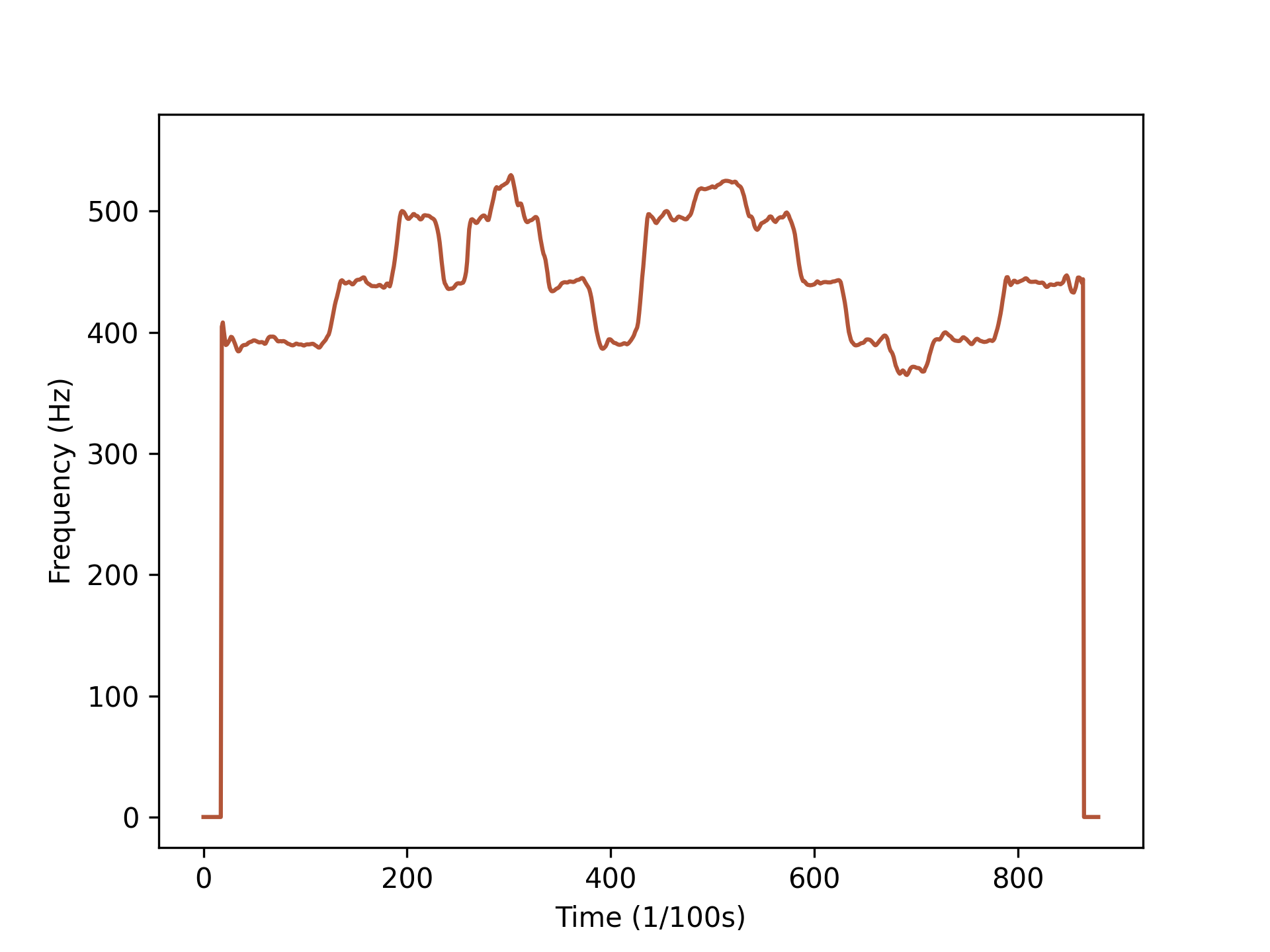}
    \caption{We first extract the $f_0$ using RVMPE.}
\end{subfigure}\hspace{\fill} 
\begin{subfigure}[t]{0.45\linewidth}
    \includegraphics[width=\linewidth]{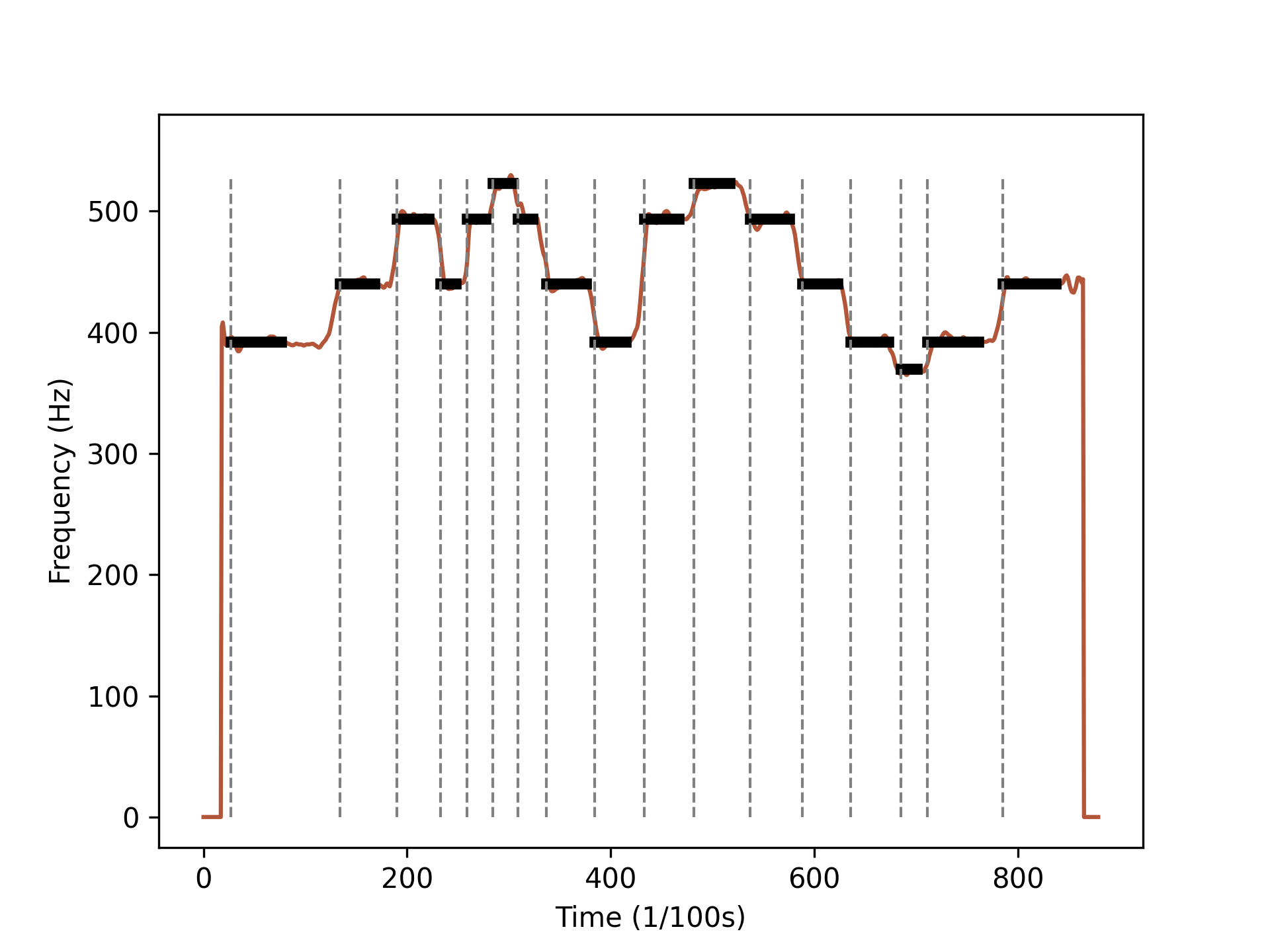}
    \caption{Based on our MIDI data computed by Basic Pitch, we segment the $f_0$ curve by looking at each note onset.}
\end{subfigure}

\bigskip 
\begin{subfigure}[t]{0.45\linewidth}
    \includegraphics[width=\linewidth]{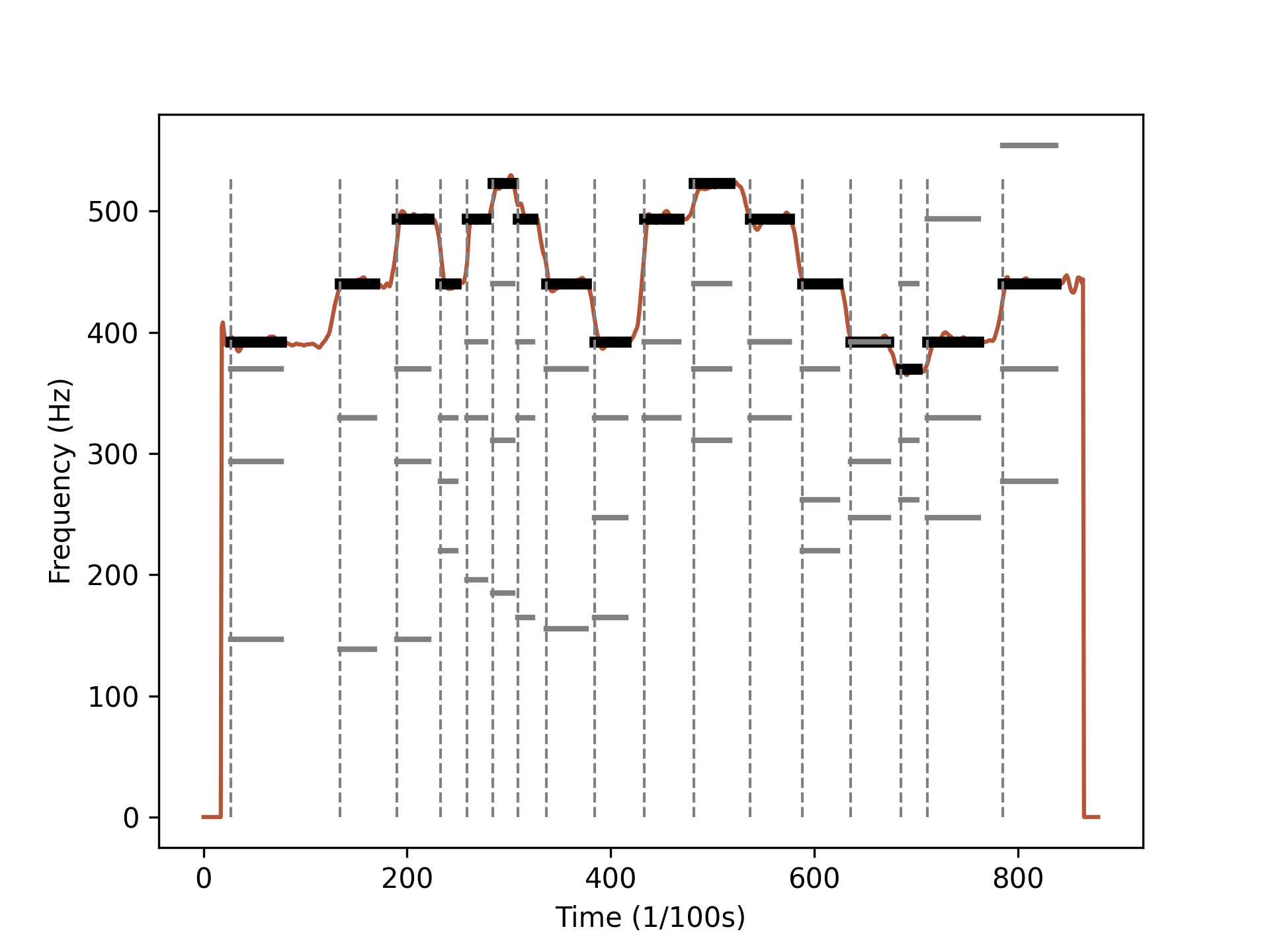}
    \caption{We obtain a four-part harmony from the Anticipatory Music Transformer.}
\end{subfigure}\hspace{\fill} 
\begin{subfigure}[t]{0.45\linewidth}
    \includegraphics[width=\linewidth]{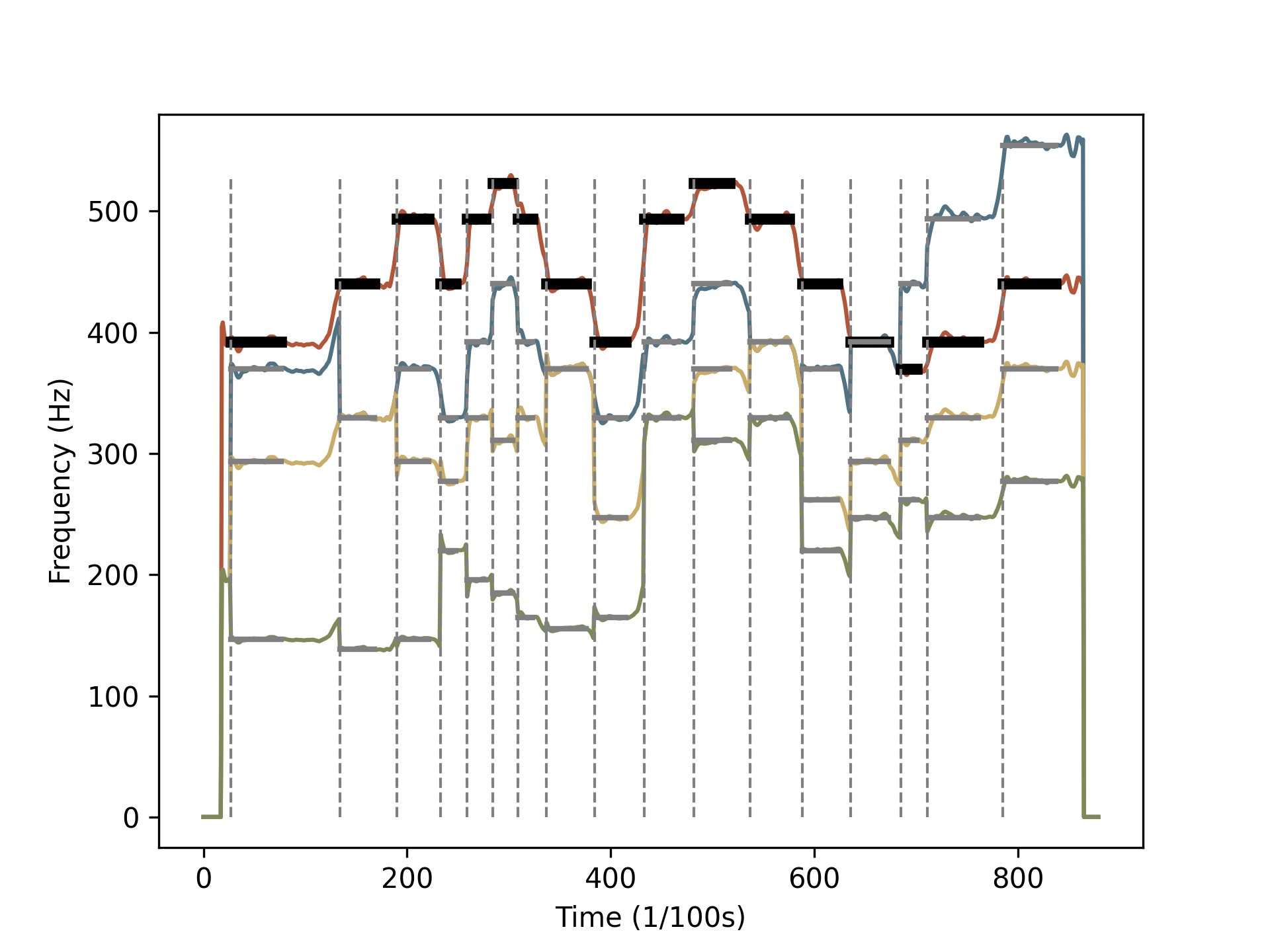}
    \caption{We shift the $f_0$ curve for each note of every voice, and obtain four distinct curves.}
\end{subfigure}
\caption{Segmentation and $f_0$ shifting process for our automatic harmonization.}
\label{fig:f0}
\end{figure}

An overview of our pitch shifting approach is presented on Figure \ref{fig:f0}.

\subsection{Voice Synthesis with RVC}

We finally perform voice synthesis using the new $f_0$ curves calculated as described above for each harmonic voice. For this step, RVC requires us to use a pre-trained vocal model that contains an index file of HuBERT embeddings, as well as weights for the feature encoder, normalizing flow, and HiFi-GAN vocoder as shown in Figure \ref{fig:arch}. This requires us to train the user's voice model beforehand. Once this model is trained, we can use both the modified $f_0^{out}$ curve for each voice as well as the extracted HuBERT embeddings for the input audio to condition the synthesis of each vocal line. With this pitch conditioning, RVC allows us to preserve the formant and timbre of the original audio, while following the provided pitch information, thus producing natural-sounding harmonies that maintain the characteristics of the original singer.

Once the audio is generated for every additional vocal line, we can add them together and retrieve the final harmonized vocal line.

\section{Results \& Discussion}

We tested our system on a variety of audio inputs and obtained highly convincing results. While the system demonstrates significant power, its complexity poses challenges for real-time adaptation as a musical instrument. To facilitate future research on adapting similar systems into real-time performance settings, we provide the results of our experiments, focusing on inference speed across different hardware configurations.

\begin{figure}

\begin{subfigure}[t]{\linewidth}
    \includegraphics[width=\textwidth]{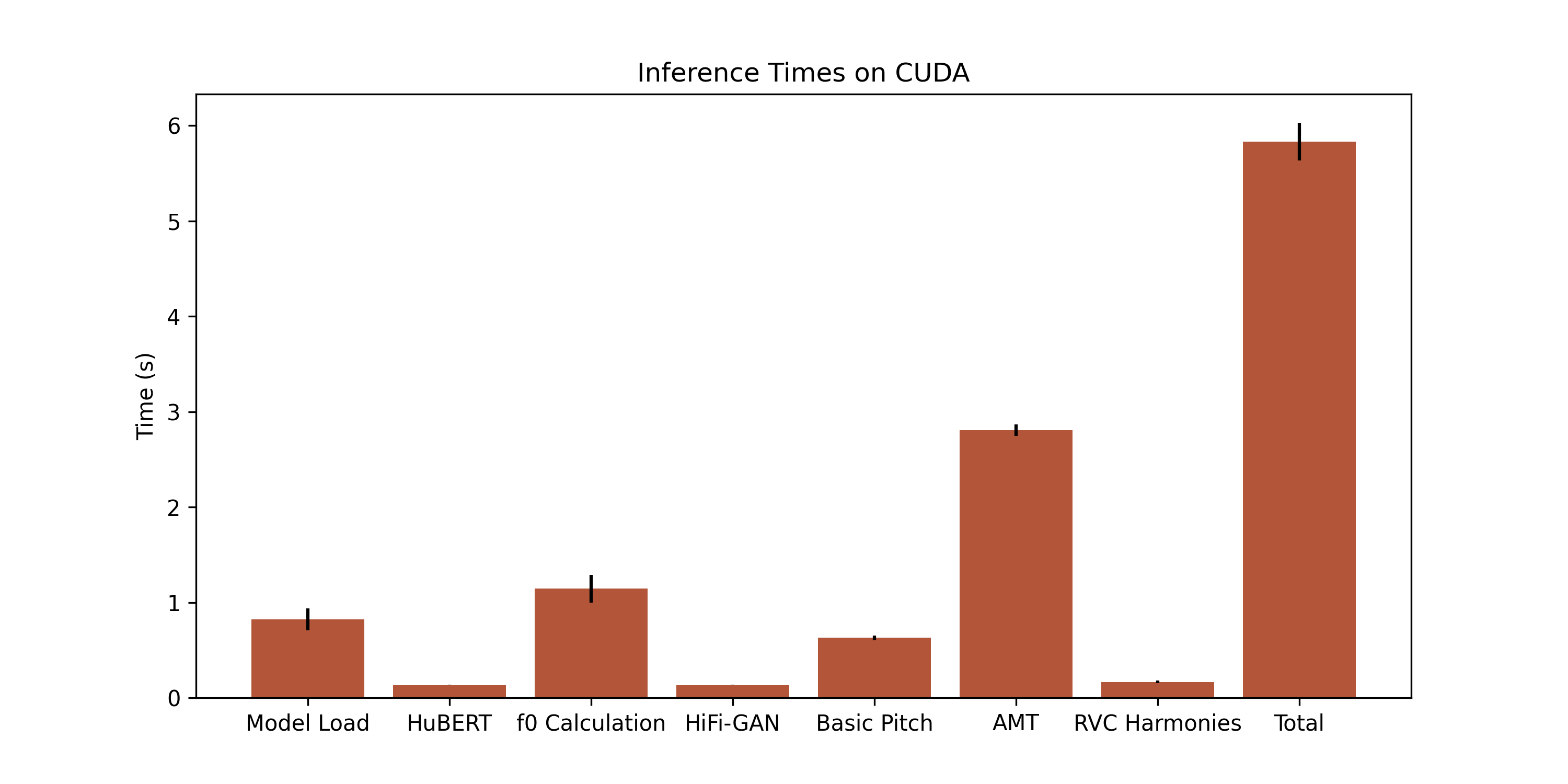}
\end{subfigure}\hspace{\fill} 
\begin{subfigure}[t]{\linewidth}
    \includegraphics[width=\linewidth]{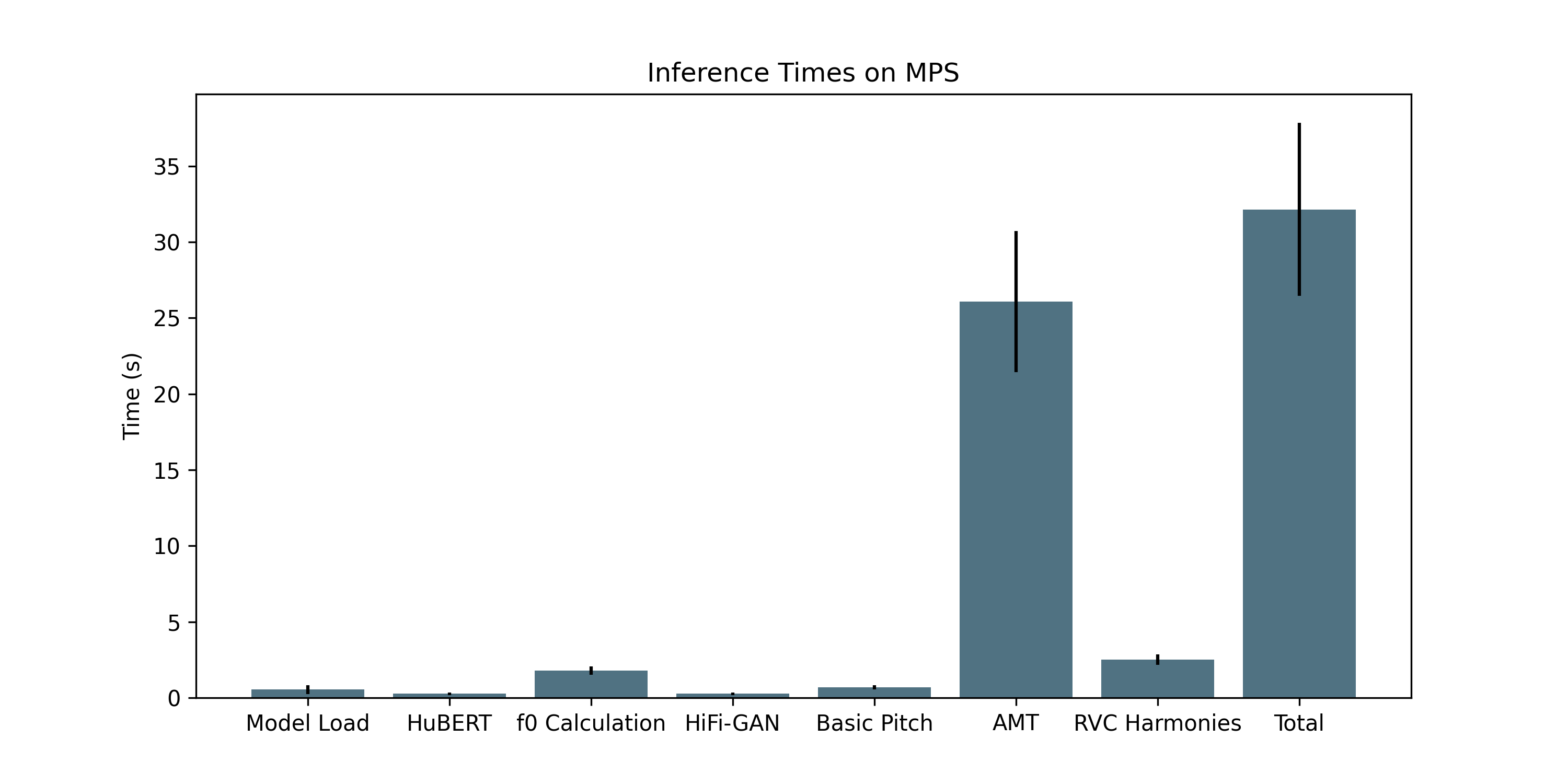}
\end{subfigure}

\caption{Comparison of inference times on machines running CUDA and MPS, with a 10-second audio input.}
\label{fig:inference}
\end{figure}

Our tests were conducted on two machines: An RTX 4090 GPU Machine running Ubuntu 22.04 and a M3 Max MacBook Pro running macOS 14.5. Figure \ref{fig:inference} presents the inference time of each system component. Notably, our results indicate that inference on the CUDA-powered machine is significantly faster than on the MPS-based MacBook. This discrepancy appears to stem from a known issue in the PyTorch library, where iterative inferences--essential for autoregressive models like AMT--lead to substantial memory leaks, ultimately causing performance slowdowns.\footnote{See the open PyTorch issue on GitHub: \url{https://github.com/pytorch/pytorch/issues/91368}.}

Despite this, our findings are encouraging: on the CUDA system, a full iteration of the model for a 10-second audio input completes in under six seconds on average. Achieving similar performance on the MacBook Pro may be possible by optimizing AMT inference times. The second most time-consuming step is the $f_0$ calculation, which could likely be accelerated by replacing RMVPE with a more efficient model, like PESTO \cite{PESTO}. Further improvements could be gained by training a unified model capable of simultaneously predicting both the $f_0$ and a MIDI transcription from the audio input.

Looking ahead, we envision this technology playing a crucial role in real-time musical performance. While further optimizations are necessary, real-time pitch tracking systems and improvements to the Music Transformer architecture could enable harmonization at speeds faster than real-time.

\section{Conclusion}

In this paper, we introduced the AI Harmonizer, a novel system capable of autonomously generating four-part vocal harmonies without user-provided harmonic input. By leveraging state-of-the-art AI models, our framework successfully arranges input melodies into rich choral textures. Our experimental results demonstrate the effectiveness of our approach in producing musically coherent harmonies that preserve the vocal characteristics of the original singer. Additionally, our performance analysis across different hardware configurations highlights the system's potential for real-time application, particularly with optimizations in pitch tracking and inference processes. Despite the system currently operating offline, these advancements represent a significant step toward AI-assisted vocal performance and expressive musical augmentation. Future work will focus on refining the harmonization process for broader musical contexts beyond SATB chorales and exploring interactive performance applications. We also intend to investigate the potential for machine learning models that further bridge the gap between symbolic and audio representations. By continuing to push the boundaries of AI in music, we hope to empower artists with innovative tools that expand creative possibilities in both composition and live performance.

\section{Ethical Standards}

There are no observed conflicts of interest. The research was funded using discretionary funding and used lab-owned compute power for the training of the model. Consent by the vocal performer was obtained before training a voice model and distributing the vocal audio recordings.

\section*{Acknowledgments}

We are grateful to Nancy Rosenberg for providing the vocal samples we used to train and test our system.

\bibliographystyle{ACM-Reference-Format}
\bibliography{references}

\end{document}